\documentclass[twocolumn,prl,superscriptaddress,floats,nobibnotes,notitlepage,citeautoscript]{revtex4-2}
\usepackage{graphicx}
\usepackage{hyperref}
\usepackage{epstopdf}
\usepackage[utf8]{inputenc}
\usepackage{amsmath,bm}
\usepackage{bbold}
\usepackage{color,soul}
\usepackage{xstring}
\usepackage{wasysym}
\usepackage{xspace}
\usepackage{amssymb}
\usepackage{time}
\usepackage{bbold}
\usepackage{subfigure}
\usepackage{multirow}
\usepackage{xspace}
\usepackage{url}
\usepackage{amsmath}
\usepackage{amssymb}
\usepackage{ulem}
\usepackage[dvipsnames]{xcolor}
\usepackage{orcidlink}

\def\<{\langle}
\def\>{\rangle}

\newcommand{\ve}[1]{\boldsymbol{#1}}

\begin{document}

\title{
Edge modes of topological Mott insulators and deconfined  quantum critical points}

\author{\firstname{Yuhai} \surname{Liu}}
\email{yuhailiu@bupt.edu.cn}
\affiliation{\mbox{School of Physical Science and Technology, Beijing University of Posts and Telecommunications, Beijing 100876, China}}
\affiliation{\mbox{Beijing Computational Science Research Center, Beijing 100193, China}}
\author{ Toshihiro Sato}
\affiliation{\mbox{Institute for Theoretical Solid State Physics, IFW Dresden, 01069 Dresden, Germany}}
\affiliation{\mbox{W\"urzburg-Dresden Cluster of Excellence ct.qmat, Germany}}
\author{Disha Hou}
\affiliation{\mbox{Institut f\"ur Theoretische Physik und Astrophysik, Universit\"at W\"urzburg, 97074 W\"urzburg, Germany}}
\affiliation{\mbox{W\"urzburg-Dresden Cluster of Excellence ct.qmat, Germany}}
\author{Zhenjiu Wang }
\email{wangzj@lzu.edu.cn}
\affiliation{Lanzhou Center for Theoretical Physics, Key Laboratory of Quantum Theory and Applications of MoE, Key Laboratory of Theoretical Physics of Gansu Province, and School of Physical Science and Technology, Lanzhou University, Lanzhou, Gansu 730000, China.}
\affiliation{Arnold Sommerfeld Center for Theoretical Physics, University of Munich, Theresienstr. 37, 80333 Munich, Germany}
\affiliation{Max Planck Institute for the Physics of Complex Systems, N\"othnitzer Strasse 38, Dresden 01187, Germany}
\author{\firstname{Wenan} \surname{Guo}}
\email{waguo@bnu.edu.cn}
\affiliation{School of Physics and Astronomy, Beijing Normal University, Beijing 100875, China}
\affiliation{Key Laboratory of Multiscale Spin Physics (Ministry of Education), Beijing Normal University, Beijing 100875, China}
\author{Fakher F. Assaad\, \orcidlink{0000-0002-3302-9243}}
\email{assaad@physik.uni-wuerzburg.de}
\affiliation{\mbox{Institut f\"ur Theoretische Physik und Astrophysik, Universit\"at W\"urzburg, 97074 W\"urzburg, Germany}}
\affiliation{\mbox{W\"urzburg-Dresden Cluster of Excellence ct.qmat, Germany}}

\begin{abstract}
Topology and anomalies lead to edge modes that can interact with critical bulk fluctuations.  To study
this setup, pertaining to boundary  criticality, we consider a model exhibiting a deconfined quantum critical point (DQCP)
between a dynamically generated quantum spin Hall state—a topological Mott insulator—and an s-wave superconductor. For the topological Mott insulator,
the bulk Goldstone modes are shown to be irrelevant at the helical Luttinger liquid fixed points. The deconfined quantum critical point is an instance of an emergent anomaly,
and we observe a sharp localized edge state at this point. The sharpness of the edge mode is consistent with an ordinary phase in which electronic edge
modes decouple from critical edge bosonic fluctuations.
At the DQCP, the scaling dimension of the edge electron shows a jump, a feature argued to be a signature of the emergent anomaly.
Our results are based on large-scale auxiliary-field quantum Monte Carlo simulations. We also carry out calculations for the Kane-Mele-Hubbard model to confirm spectral
features of the ordinary and extraordinary-log phases in the vicinity of the bulk critical point.
\end{abstract}

\maketitle

 \textit{Introduction}--- Topological insulators are characterized by a bulk gap with symmetry protected edge modes on open manifolds \cite{Moore10}. Canonical examples are the Haldane model
 with chiral edge modes \cite{Haldane98}. Edge states are a consequence of the bulk topology: gauging the theory and integrating out the fermions leads to a Chern-Simons action for the vector potential. It encodes the distinct electromagnetic response of  Chern insulators and the necessity of edge states on open manifolds due to charge conservation  \cite{Witten16, Qi08a, HoAsreview2013}. A  single edge state of a topological insulator can be a realization of an anomaly: an isolated chiral edge state of the Haldane insulator cannot be realized by a local Hamiltonian \cite{Nielsen81, WangZ22}. This anomaly is matched by the edge state on the other side of the sample, such that local Hamiltonians containing no \textit{net} anomaly are possible to
 realize.

Bulk anomalies provide a distinct route to edge states; edge modes are required to match the bulk anomalies.
The distinction from topological insulators lies in the fact that,  in this case, the bulk is gapless \cite{Thorngren20}.  A canonical example is Weyl semi-metals \cite{Wan83}  with surface  Fermi arcs that are protected by the chiral anomaly.   Other examples include one dimensional versions of  deconfined quantum criticality  between  bond density  and  charge density wave orders \cite{yang2025dqcp-1d-gapless} where the single
particle gap remains open across the transition.  In fact this critical point
corresponds to an intrinsically  gapless symmetry protect state as coined by the authors of Ref.~\cite{Thorngren20}.

Generically, defects such as boundaries lead to new exponents \cite{Cardy96_Book,Toldin17}.
Even in a mean-field approximation of the 3D Ising model with identical bulk and edge couplings, surface exponents differ from the bulk \cite{Cardy96_Book}, corresponding to the so-called ordinary phase. Upon enhancing the edge coupling, the surface will order while the bulk remains critical. This is the so-called extraordinary phase \cite{Cardy96_Book}.
In 3D O($N \ge 2$) models, new phases have recently been discovered at the boundary \cite{Metlitski20, Hu2021, ParisenToldin2021, Toldin22}, the extraordinary-log phase \cite{Metlitski20, Hu2021, ParisenToldin2021, Toldin22}  where correlation functions on the surface decay with a power of the logarithm of the distance.

Topology and anomalies lead to new degrees of freedom at the edge that can interact with the  boundary operators of the bulk critical phenomena \cite{Grover12}.
The question we address in this article is the fate of the edge state when the bulk has scale-invariant fluctuations or is in the proximity of a quantum critical point.
Recently, the authors of Ref.~\cite{Toldin25} have studied the Kane-Mele-Hubbard model \cite{Hohenadler10, Hohenadler12} with helical Luttinger liquid edge states tuned to the 3D XY bulk critical point.
An extraordinary-log (ordinary) phase has been numerically observed when the aforementioned coupling between the edge mode and order parameter fluctuations is relevant (irrelevant).
Similar calculations have been carried out for the Kane-Mele Hubbard model in the presence of spin-orbit coupling that reduces the 3D XY bulk criticality to $\mathbb{Z}_2$ using numerical \cite{Ge25} and analytical
methods  \cite{Shao-Kai2025}.

\begin{figure}[t]
  \centering
  \centerline{\includegraphics[width=0.5\textwidth]{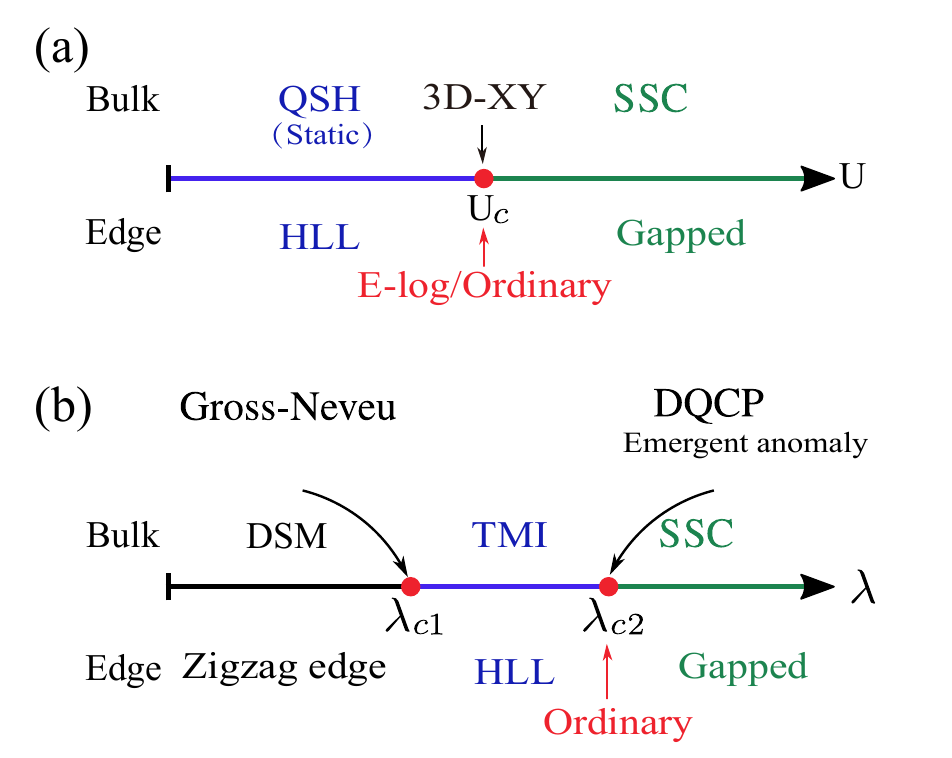}}
    \caption{\label{fig:Phase_diagram}
  (a) Bulk phase diagram of the Kane-Mele-Hubbard with spin-orbit coupling $\lambda_{KM}=0.1$~depicts the occurrence of static QSH and SSC phases.
 Depending upon the value of the boundary Hubbard term,  ordinary and extraordinary phases are realized at the edge \cite{Toldin25}.
  (b) Schematic ground-state phase diagram in the interaction strength $\lambda$ with Dirac semi-metal
  (DSM), dynamic quantum spin Hall (QSH), and s-wave superconductor (SSC) phases for the bulk, as presented in Ref.~\cite{Liu18}.
  }
  \end{figure}

We consider three case scenarios and two models,  the phase diagrams of which are depicted in Fig.~\ref{fig:Phase_diagram}.
In contrast to Ref.~\cite{Toldin25}, we will vary the bulk coupling to tune through the bulk quantum phase transition.
To set the stage, we start with the Kane-Mele model \cite{KaneMele05}   supplemented by an attractive  Hubbard interaction \cite{Hohenadler10} (See Fig.~\ref{fig:Phase_diagram}a). As documented in Ref.~\cite{Hohenadler12},  the Hubbard interaction drives the bulk through
a 3D XY transition. The fate of the edge state at the bulk critical point has recently been studied in great detail in Ref.~\cite{Toldin25}, the outcome being a realization of an extraordinary-log phase in this quantum model. Next, we will consider a model introduced in Ref.~\cite{Liu18}, that supplements the Hamiltonian of Dirac fermions on the honeycomb lattice with the square of a \textit{local} Kane-Mele spin-orbit interaction term.  This fermionic model has  SU(2) spin symmetry,  U(1) charge symmetry,  and a phase diagram depicted in Fig.~\ref{fig:Phase_diagram}(b). We observe a Gross-Neveu  (GN) transition from a  Dirac semi-metal (DSM)  to a dynamically generated quantum spin Hall (QSH) state.  This state is a realization of a topological Mott insulator (TMI)\cite{Raghu08} with scale-invariant Goldstone modes in the bulk that may
interact with the edge state. The quantum critical point between the  QSH  and s-wave superconducting (SSC) phase is an instance of deconfined quantum criticality \cite{Senthil04_1, Senthil04_2} (DQC).   At this point, the single particle excitations remain gapped such that the symmetry of the model is  SO(3) $ \times $ U(1). This symmetry group is anomalous \cite{Thorngren20} such that one expects distinct fermionic edge modes that cancel the anomaly at this critical point  \cite{Metlitski18, Thorngren20}.

\textit{ Model and Method}
\label{sec:Model}--- The attractive Kane-Mele-Hubbard (KMH) Hamiltonian is defined on the honeycomb lattice. It reads,
\begin{equation}\label{Eq:KHM1}
\begin{aligned}
  \hat{H} &  = - t  \sum_{ \langle \bm{i}, \bm {j} \rangle }  \left(\hat{\ve{c}}^{\dagger}_{\bm{i} } \hat{\ve{c}}^{\phantom\dagger}_{\bm{j}}    + \text{H.c.} \right)
-\lambda_{ \text{KM} } \sum_{\varhexagon}  \left( \sum_{\langle \langle \bm{i}, \bm{j} \rangle \rangle  \in \varhexagon }
   \hat{  {J}}_{\bm{i},\bm{j}}^z \right)  \\
 &- \frac{U}{2} \sum_{\bm i} (  \hat n_{ \bm i, \uparrow } + \hat n_{ \bm i, \downarrow } -1 )^2
\end{aligned}
\end{equation}
where the spinor
$\hat{\boldsymbol{c}}^{\dag}_{\ve{i}} =
\big(\hat{c}^{\dag}_{\ve{i},\uparrow},\hat{c}^{\dag}_{\ve{i},\downarrow}
\big)$, $\hat{c}^{\dag}_{\ve{i},\sigma} $ creates an electron at lattice
site $\ve{i}$ with $z$-component of spin $\sigma$, $\hat{n}^{\phantom\dagger}_{\ve{i},\sigma} = \hat{c}^{\dag}_{\ve{i},\sigma} \hat{c}^{\phantom\dagger}_{\ve{i},\sigma}$.
$ \langle \bm{i}, \bm{j} \rangle $  ($\langle \langle \bm{i}, \bm{j} \rangle \rangle $) denote
(next)  nearest neighbor sites,
$\hat{J}^z_{\bm{i},\bm{j}}  =  i \nu_{ \bm{i} \bm{j} }
 \hat{\ve{c}}^{\dagger}_{\bm{i}} \sigma^z
\hat{\ve{c}}^{\phantom\dagger}_{\bm{j}}+\rm{H.c.}$  with
$\nu_{\boldsymbol{ij}}= - \nu_{\boldsymbol{ji}} = \pm1$.  For the phase factors, $\nu_{ \bm{i} \bm{j} }$,  we  use the same notation as in \cite{KaneMele05}: let  $\ve{r}$  be the nearest neighbor site
common to  next nearest neighbor  sites $\ve{i}$ and  $\ve{j} $, then
$	   \nu_{\ve{ij}} =       \text{ sgn } \left[  \left( \ve{i}  - \ve{r} \right)  \times  \left( \ve{r}  - \ve{j} \right)  \right]    \cdot  \ve{e}_{\perp}$, with $\ve{e}_{\perp}$ a vector perpendicular to the Honeycomb plane. We set $t=1$ and $\lambda_{ \text{KM} } = 0.1$ such that the bulk single particle gap of the  QSH insulator at  $U=0$ is set by  $0.3\sqrt{3}$. For this parameter choice, the bulk 3D XY phase transition depicted in Fig.~\ref{fig:Phase_diagram}(a) between a static QSH and SSC phases is known to be $ U_c \approx 4.9$ \cite{Hohenadler10, Hohenadler12}.

The $t-\lambda$ model is again defined on the honeycomb lattice  and reads:
\begin{eqnarray}\label{Eq:Ham}
 \hat{H}  & = &- t  \sum_{ \langle \bm{i}, \bm {j} \rangle }  \left( \hat{\ve{c}}^{\dagger}_{\bm{i} } \hat{\ve{c}}^{\phantom\dagger}_{\bm{j}} + \text{H.c.} \right)
   -\lambda \sum_{\varhexagon}  \left( \sum_{\langle \langle \bm{i} \bm{j} \rangle \rangle  \in \varhexagon }   \hat{\ve{J}}_{\bm{i},\bm{j}} \right)^2
\end{eqnarray}
where
$\hat{\ve{J}}_{\bm{i},\bm{j}}  =  i \nu_{ \bm{i} \bm{j} }
 \hat{\ve{c}}^{\dagger}_{\bm{i}} \bm{\sigma}
\hat{\ve{c}}^{\phantom\dagger}_{\bm{j}}+\rm{H.c.}$ with $\boldsymbol{\sigma}=(\sigma^x,\sigma^y,\sigma^z)$ the vector of  Pauli
spin matrices.
Since  $ \hat{\ve{J}}_{\bm{i},\bm{j}} $  transforms as a vector under   SU(2) spin rotation,  the model possesses global SU(2) spin rotation symmetry.

We have investigated the two model Hamiltonians (\ref{Eq:KHM1}) and (\ref{Eq:Ham})  using the  ALF (Algorithms for Lattice Fermions) implementation~\cite{ALF_v1,alfcollaboration2021alf}  of the  well-established finite temperature auxiliary-field quantum Monte Carlo (AFQMC) method~\cite{Blankenbecler81, Hirsch85, White89, Assaad08_rev}. Since both interactions are written in terms of squares of single-body operators, the model is readily implemented in the ALF-library. We consider values of $U>0$
for the KMH model and $\lambda>0$  for the $t-\lambda$ model  such that for a given instance of Hubbard-Stratonovitch fields, both models preserve time reversal symmetry which ensures that the eigenvalues of the fermionic matrix occur in complex conjugate pairs  \cite{Wu04}. Hence, no negative sign problem occurs. We have used a Trotter discretization $\Delta_{\tau}= 0.2$ and a symmetric Trotter decomposition to ensure the hermiticity of the imaginary time propagation. To study the edge physics, we consider periodic boundary conditions only in one direction of the two-dimensional lattice and open boundaries in the other one. Specifically we consider a one-dimensional lattice  with lattice vector $\bm{a}_x$  and  periodic  boundary conditions with unit cell containing $N_{orb}$ orbitals.  The cut is such that we have zigzag edges (see  Fig.~\ref{fig:lattice_diagram} in the End Matter).  We simulated lattices with $L_x$ unit cells (each containing $N_{orb}=L_x$ orbitals) and set the inverse  temperature to $\beta = 2L_x$.  Unless mentioned otherwise bulk and boundary
couplings are identical.

{\it QMC Results.}\label{sec:Result}---
\begin{figure}[t]
  \centering
\includegraphics[width=0.45\textwidth]{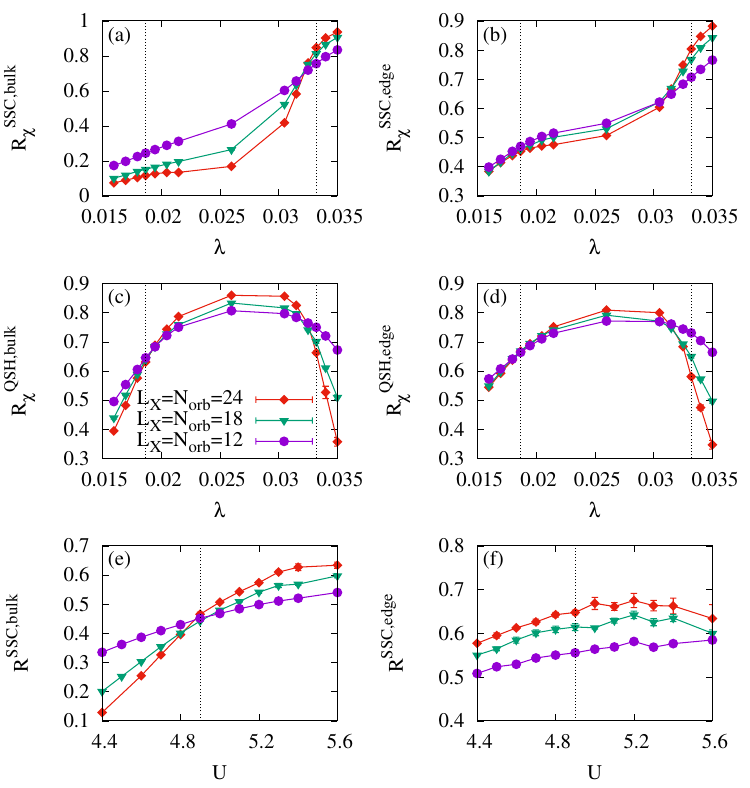}
  \caption{\label{fig:QSH_SSC_Ratio}
  $\lambda$ dependence of time displaced correlation ratios for SSC [(a)-(b)] and QSH [(c)-(d)] orders in the bulk (left panel) and along the edge (right panel) for the $t-\lambda$ model of Eq. ~(\ref{Eq:Ham}). Hubbard $U$ dependence of  equal-time   correlation ratio for SSC order in the bulk (e) and along the edge (f) for the KMH model with spin-orbit
coupling  $\lambda_{KM}=0.1$. The correlation ratios of (a),(b),(c),(d)  ((e),(f))  stem from  susceptibilities (equal-time correlations).}
\end{figure}
To map out the ground-state phase diagram both in the bulk and on the edge, we calculate the renormalization-group invariant correlation ratio
\begin{eqnarray}
R^{O,n} =  1 - \frac{ \chi^{O,n} (k_x^O+\delta k_x)}{   \chi^{O,n} (k_x^O) }
\end{eqnarray}
using corresponding QSH and SSC susceptibilities for orbital $n$ on the boundary or in the bulk.  These quantities are defined in Eqs.~(\ref{Eq.QSH}) and (\ref{Eq.SSC}) in the End Matter.
In both cases, the ordering wave vector corresponds to $k_x^O = 0$ and $\delta k_x = 2\pi/L$.
By definition, for $L \rightarrow \infty$, $R^{O,n}\rightarrow1 (0)$ in the ordered (disordered) phase.
At the critical point, $R^{O,n}$ is scale-invariant for sufficiently large $L$, thus observing a crossing in $R^{O,n}$ for different system sizes.

In Figs.~\ref{fig:QSH_SSC_Ratio}(a), \ref{fig:QSH_SSC_Ratio}(c), and \ref{fig:QSH_SSC_Ratio}(e), we plot the correlation ratios for the bulk,  and in Figs.~\ref{fig:QSH_SSC_Ratio}(b), \ref{fig:QSH_SSC_Ratio}(d), and \ref{fig:QSH_SSC_Ratio}(f)  for the boundary. For the  $t-\lambda$ model, Figs.~\ref{fig:QSH_SSC_Ratio}(a)-\ref{fig:QSH_SSC_Ratio}(d),  we  observe bulk and boundary transitions at identical values of the coupling $\lambda$.  The dashed vertical lines denote the bulk transition point after extrapolation as obtained from Ref.~\cite{Liu18}.
For the  KMH model Fig.~\ref{fig:QSH_SSC_Ratio}(e) and \ref{fig:QSH_SSC_Ratio}(f), the situation is  different:  The bulk transition  between the QSH  and  the SSC, Fig.~\ref{fig:QSH_SSC_Ratio}(e), does  not show up on the edge. In particular, on our finite-sized system, the edge remains \textit{ordered} at $U < U_c$ where the bulk is disordered
(see Fig.~\ref{fig:QSH_SSC_Ratio}(f)).

We now proceed to examine the evolution of electronic structure in more detail by measuring the momentum resolved single-particle spectrum.
Taking into account the one-dimensional system with orbitals $n$ along the ${\bf a}_x$ direction (see Fig.~\ref{fig:lattice_diagram} in the End Matter),  this quantity is defined as
\begin{eqnarray}\label{Eq:spectral}
A_{ n } ( k_x, \omega)=-\frac{1}{\pi} {\rm Im} G_{n,k_x}(\omega+i0^{+}).
\end{eqnarray}
In our QMC simulations, $A_{ n } ( k_x, \omega)$ was evaluated from the imaginary-time Green's function
\begin{eqnarray}\label{Eq:GF}
G_{n,k_x}(\tau) =\frac{1}{2}\sum_{\sigma}\langle\hat{{c}}^{\dagger}_{ n,k_x,\sigma}(\tau)\hat{{c}}^{\phantom\dagger}_{n,k_x,\sigma}(0)\rangle
\end{eqnarray}
using the ALF \cite{alfcollaboration2021alf} implementation of the  stochastic analytical continuation method~\cite{Beach04a,Sandvik98}.
Henceforth, we show the results solely for $n=1$ along the edge.

\begin{figure}[t]
  \centering
\includegraphics[width=0.48\textwidth]{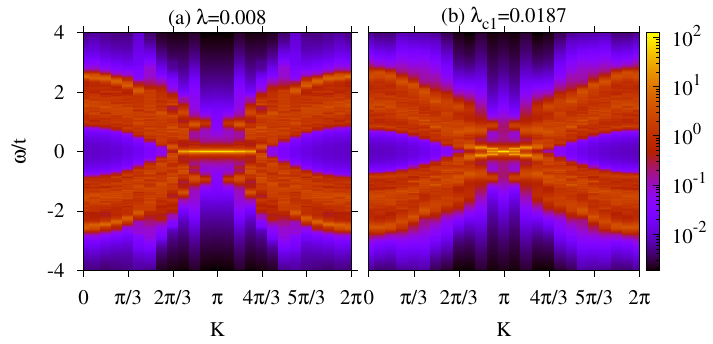}
  \caption{\label{fig:SP_GN}
  Single-particle spectral function $A_{ n } ( k_x, \omega)$ along the zigzag edge,
  inside the bulk DSM phase (a) ($\lambda=0.008$)  and at the bulk GN phase transition point (b) ($\lambda_{c1}=0.0187$).}
\end{figure}

\begin{figure}[t]
  \centering
\includegraphics[width=0.48\textwidth]{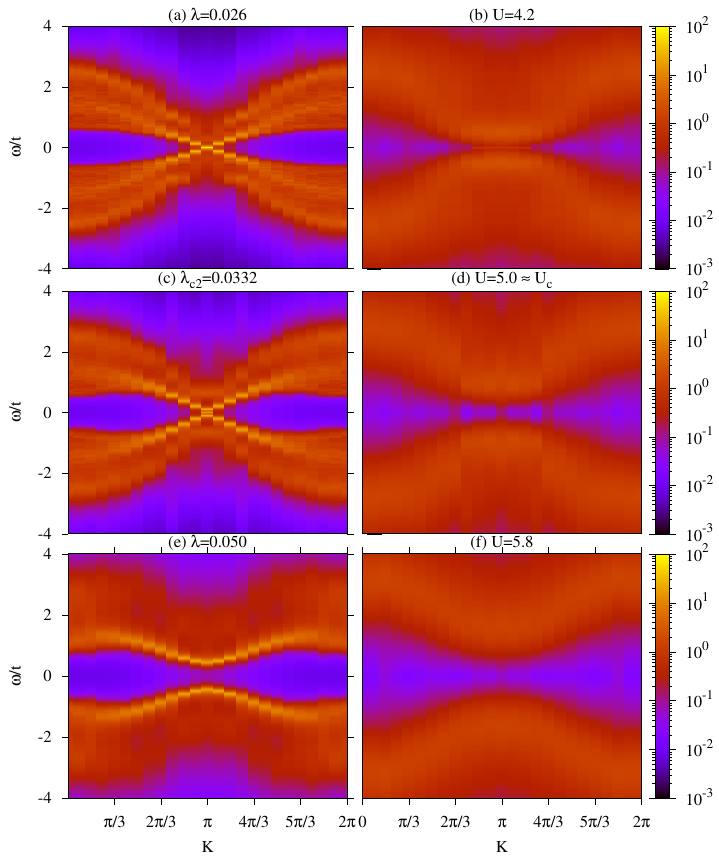}
  \caption{\label{fig:SPG_QSH_SSC}
  The single particle spectrum across the QSH-SSC phase transition for $t-\lambda$ Model [(a), (c), (e)] and across the  3D-XY transition for the Kane Mele Hubbard model [(b), (d), (f)].  }
\end{figure}

The QSH state in the $t-\lambda$ model  is dynamically  generated.  In Fig.~\ref{fig:SP_GN} we show the evolution  of  edge spectral  function across  the GN transition. The flat  zigzag  edge  state  with  gapless excitations  at  the Dirac  points,  evolves  into  a  helical  Luttinger liquid  with gapless excitations at  the  time  reversal symmetric  point $k_x = \pi$ \footnote{In this particular  case, particle-hole symmetry pins   the  Dirac  point at  $k=\pi$  to the Fermi energy.}.   The low energy degrees  of this phase correspond  to  helical  edge  states  and   Goldstone modes  of  the  spontaneously  broken  SO(3) symmetry \cite{Hohenadler21}.  Assume  symmetry  breaking  along  the z-direction  such  that the action of  the  helical liquid reads
$ S   =   \ve{c}^{\dagger}(x,\tau) \left( \partial_\tau  +  v_F  \sigma^{z} i \partial_x \right) \ve{c}^{}(x,\tau) $  with  $\ve{c}^{\dagger}(x,\tau)  =
\left( {c}^{\dagger}_{\uparrow}(x,\tau),{c}^{\dagger}_{\downarrow}(x,\tau)  \right) $  a  spinor of Grassmann  variables.  Let  $\ve{n}(\ve{x},\tau) $  encode  transverse   fluctuations  of the   QSH  order  parameter.  Importantly $\ve{n}(\ve{x},\tau) $  is   even under time reversal  symmetry such that the  Yukawa   coupling  between the edge  state  and  the helical liquid Goldstone modes,
$\ve{n}(x,\tau) \cdot \ve{c}^{\dagger}(x,\tau)  \left( \sigma^x,\sigma^y \right)  \ve{c}^{}(x,\tau) $  is   symmetry forbidden.      Allowed  couplings  include higher order terms    such as
 $(\ve{n} \cdot \partial_{\mu} \ve{n} )( \ve{c}^{\dagger} \partial_{\mu} \ve{c}^{}) $,   that   turn out to be irrelevant \footnote{ At the decoupled critical point of bulk and boundary the scaling dimension of this
operator scales as   $2 -2 - 2\Delta_n - 2 \Delta_c < 0 $  where  $\Delta_n$ and $\Delta_c$ are the scaling dimensions of the order parameter and the edge state fermion respectively.}. The  irrelevance  of the coupling  between the edge  state and  the  scale invariant  bulk   QSH  fluctuations  provides an understanding of  why the edge state  deep in the QSH  phase  at  $ \lambda =  0.026$, is  so visible in the spectral function, Fig.~\ref{fig:SPG_QSH_SSC}(a).

To study the coupling of  the  edge state to superconducting fluctuations we  adopt a  Bogoliubov  basis,  $\ve{\hat{\gamma}}^{\dagger}_{k_x}    =
\left(\ve{c}^{\dagger}_{k_x,\uparrow}, \ve{c}^{}_{-k_x,\downarrow}  \right) $    with the edge  Hamiltonian
 $\hat{H}    =  - v_F \sum_{k_x}  k_x \ve{\hat{\gamma}}^{\dagger}_{k_x} \tau^z  \ve{\hat{\gamma}}^{}_{k_x} $.
 In this basis, the  generator  of  U(1)   charge conservation  corresponds to $\tau^z$ and  the  two-dimensional  vector   consisting of the real and imaginary parts  of  the  superconducting  fluctuations  $\hat{\ve{\Delta}}$  transforms as an SO(2)  vector.    Yukawa terms of the form  $\ve{\hat{\gamma}}^{\dagger}_{x} \left(\tau^x,\tau^y\right) \ve{\hat{\gamma}}^{}_{x}  \cdot \ve{\hat{\Delta}}_{x} $   are allowed  by  symmetry and in the superconducting phase, the edge state is gapped out as seen in Fig.~\ref{fig:SPG_QSH_SSC}(e).

The data for the KMH model  shows no clear sign of a  helical Luttinger liquid (see Figs.~\ref{fig:SPG_QSH_SSC}(b), \ref{fig:SPG_QSH_SSC}(d), and \ref{fig:SPG_QSH_SSC}(f)). At the bulk critical point  and  for identical values of bulk and boundary Hubbard interactions,  this stands in agreement with Ref.~\cite{Toldin25}: The apparent pseudogap \cite{Jain24}  is a
direct consequence of the extraordinary-log phase \cite{Metlitski20}.  It is interesting to note that the features  of the extraordinary-log phase,  i) correlation functions that are hard to distinguish from genuine long-ranged order on finite-size  systems (Fig.~\ref{fig:QSH_SSC_Ratio}(f)) and  ii)  pseudogap in the
single particle spectral functions (Fig.~\ref{fig:SPG_QSH_SSC}(d)) are still seen in a wide parameter range away from bulk criticality.

\begin{figure}[t]
  \centering
\includegraphics[width=0.52\textwidth]{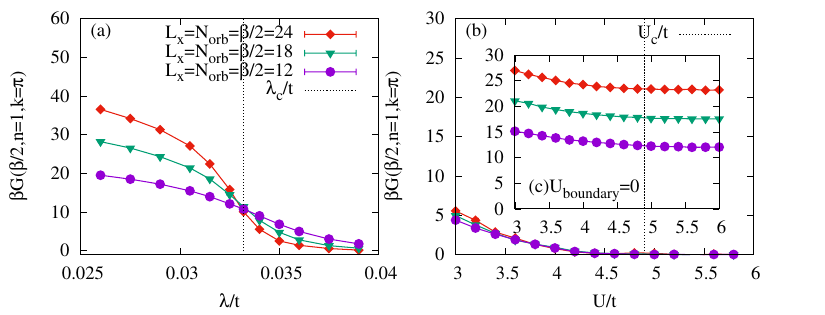}
  \caption{\label{fig:Renormalization group invariant quantity}
  $\beta G(\beta/2,n=1,k=\pi)$ across the QSH-SSC phase transition for (a),$t-\lambda$ model , (b) KMH model and (c) KMH model with $U_{\rm boundary}=0$.
 }
\end{figure}

We now  consider  the  $t-\lambda$  model  at the  DQCP.  At this point Fig.~\ref{fig:SPG_QSH_SSC}(c)  shows a pronounced edge state. As shown in Ref.~\cite{Liu18}, at the bulk critical point, the single particle gap does not close
such that the  fermion  parity $(-1)^{\sum_{\sigma} \hat{c}^{\dagger}_{\ve{i},\sigma}\hat{c}^{}_{\ve{i},\sigma} } $  is  pinned  to  unity in the infrared (IR).  Since  $2 \pi $ SU(2)  rotations  correspond  to the fermion parity, $ e^{i 2 \pi \ve{e}\cdot \ve{S}_{\ve{i}} } = (-1)^{\sum_{\sigma} \hat{c}^{\dagger}_{\ve{i},\sigma}\hat{c}^{}_{\ve{i},\sigma} }$,    they map onto unity in the  IR  limit.  Hence in the IR,  SU(2)  symmetry gives way  to SO(3).
SO(3)$\times$U(1) has a quantum anomaly in the effective field theory \cite{Thorngren20}  such that at the  DQCP a distinct   edge  state  should appear.
In Ref.~\cite{Ma21}  it has been put forward  that  the scaling  dimension of  the fermion operator shows a  discontinuity   precisely  at the DQCP.
 A  proxy for this quantity reads:
\begin{equation}
  \beta G_{1,\pi}(\beta/2) = \int d \omega  g_{\beta}(\omega)  A_1(\pi,\omega).
\end{equation}
In the  low temperature limit $g_{\beta}(\omega)$   converges  to  the  Dirac $\delta$-function   such  that the above quantity  is  a measure  of  the scaling dimension.   In fact  for  $A(\omega) \propto \frac{1}{\omega^{\alpha}}$, as appropriate for a Luttinger liquid \cite{Giamarchi}, $ \beta G_{1,\pi}(\beta/2)    \propto  \beta^{2 - \alpha} $.

To best interpret our results, we investigate the above quantity for the ordinary phase, $U_{\text{boundary}} = 0 $ and  extraordinary-log phase, $U_{\text{boundary}} = U_{\text{bulk}} $ of the KMH model.  Upon varying the
value of $U/t$, Figs.~\ref{fig:Renormalization group invariant quantity}(b) and \ref{fig:Renormalization group invariant quantity}(c)
remain featureless when tuning across the critical point. In the extraordinary-log phase,  the pseudo-gap feature leads to very small spectral weight at the Fermi energy and in the vicinity of the critical point such that $\beta G_{1,\pi}(\beta/2)$ remains very small and size independent.  In contrast, in the ordinary phase,  $\beta G_{1,\pi}(\beta/2)$ grows as a function of inverse temperature $\beta$. Using the result of  Ref.~\cite{Toldin23} for  the scaling dimension  of the boundary field operator of the 3D XY universality class, and the  scaling dimension of the pairing fluctuations of the HLL \cite{Hohenadler11a} the coupling between these two fields is irrelevant provided that   the Luttinger exponent satisfies $ K < 1.29  $. Thereby  $\alpha = 1 - (K-1)^2/2K  > 0.97$ \cite{Meden92} and is bounded by unity for attractive interactions ($K>1$).  Hence, we expect $ \beta G_{1,\pi}(\beta/2)$ to scale with $\beta$.  To a first approximation this is consistent with the data of Fig.~\ref{fig:Renormalization group invariant quantity}(c).

For the QSH-SSC  transition, $\beta G_{1,\pi}(\beta/2)$ depicted in  Fig.~\ref{fig:Renormalization group invariant quantity}(a) shows a very different behavior.
As a function of the euclidean volume, and for our lattice sizes, we observe a  divergent  behavior at $\lambda < \lambda_{c_2}$ in the QSH phase  and  a  $\beta$-independent value  at  criticality.
Furthermore, since we observe a sharp edge mode at
criticality as opposed to a pseudogap, our data supports  the notion that edge criticality of the  $t-\lambda$  model at
$\lambda = \lambda_{c_2}$ is in the ordinary phase.  A jump  in the scaling dimension of the electron as suggested by our data, has been  put  forward in Ref.~\cite{Ma21}.  The picture is based on an SU(2) parton construction of the electron in terms of bosonic and fermionic modes, the electron being composite particle of both degrees of freedom.  Precisely at the DQCP the bosonic mode is critical and  described by a $CP^{1}$ gauge theory \cite{Senthil04_2,WangC17}, whereas in the QSH phase it is condensed and the fermion mode becomes the electronic degree of freedom.  Since in the ordinary phase the fermion edge  mode decouples from the critical edge bosonic fluctuations, the scaling dimensions of electron at criticality  are given by the sum of both scaling dimensions. In contrast, in the QSH phase the scaling dimension of the electron matches that of
the fermion. This leads to a jump in the scaling dimension of the edge electron at the DQCP as observed in the QMC simulations (see Fig.~\ref{fig:Renormalization group invariant quantity}(a)).

\textit{Conclusions} --
Topology  and anomalies  lead to protected edge states.  This provides a novel area of research combining  boundary criticality, topology and anomalies.  Aside from considering the Kane-Mele Hubbard model, we have concentrated on the
t-$\lambda$ that provides realizations of a DQCP with an emergent anomaly as well as a topological Mott insulator. The t-$\lambda$ model is unique to study edge states of the DQCP since the SU(2)$\times$U(1) symmetry is not broken by the edge.  This
stands in contrast to other realizations of the DQCP such as the JQ \cite{Sandvik07}  or loop models \cite{Nahum15}  where the U(1) symmetry is emergent and will be broken by the edge.
The DQCP is known to be a weakly first order transition \cite{HeYC23,Takahashi24,Goetz24b}, but nevertheless pseudo-criticality is observed on finite lattices.
Further study of this edge state is certainly desirable.

\begin{acknowledgments}
\textit{Acknowledgments -- } We would like to thank H.Q. Lin, M. Metlitski,  F. Parisen Toldin,  S.  Ryu,  R. Thorngren, A. Vishwanath,   R. Verresen,  C. Wang  and  C. Xu  for
valuable discussions.
F.F.A.  would like to thank M. Metlitski and F. Parisen Toldin for discussions and collaborations on a related project.
The authors gratefully acknowledge the Gauss Centre for Supercomputing e.V. for funding this project by providing computing time on the GCS Supercomputer SUPERMUC-NG at Leibniz Supercomputing Centre.
D.H. and F.F.A. acknowledges the  DFG  for   funding  via  W\"urzburg-Dresden Cluster of
Excellence on Complexity and Topology in Quantum Matter
ct.qmat (EXC 2147, Project ID 390858490)  as  well  as
F.F.A. under the  grant AS 120/19-1, Project ID  530989922.
Y.L. was supported by the National Natural Science Foundation of China under Grant No.~12305039 as well as the Fundamental Research Funds for the Central Universities from the Beijing University of Posts and Telecommunications under Grant No.~2025JCTP08 and No.~2023RC42.
Z.W. was supported by the FP7/ERC Consolidator Grant QSIMCORR, No. 771891.
T.S. acknowledges funding from the Deutsche Forschungsgemeinschaft under Grant No. SA 3986/1-1 as well as the W\"urzburg-Dresden Cluster of
Excellence on Complexity and Topology in Quantum Matter
ct.qmat (EXC 2147, Project ID 390858490).
W.G. were supported by the National Natural Science Foundation of
China under Grants No. 12175015 and No. 11734002.

\end{acknowledgments}

\bibliography{./fassaad}
\clearpage
\onecolumngrid

\onecolumngrid
  \parbox[c][4em][c]{\textwidth}{\centering \large\bf End Matter}
\smallskip
\twocolumngrid

\section{Correlation functions and lattice geometry}
\label{sec:Observables}
\begin{figure}[h]
\centering
\centerline{\includegraphics[width=0.5\textwidth]{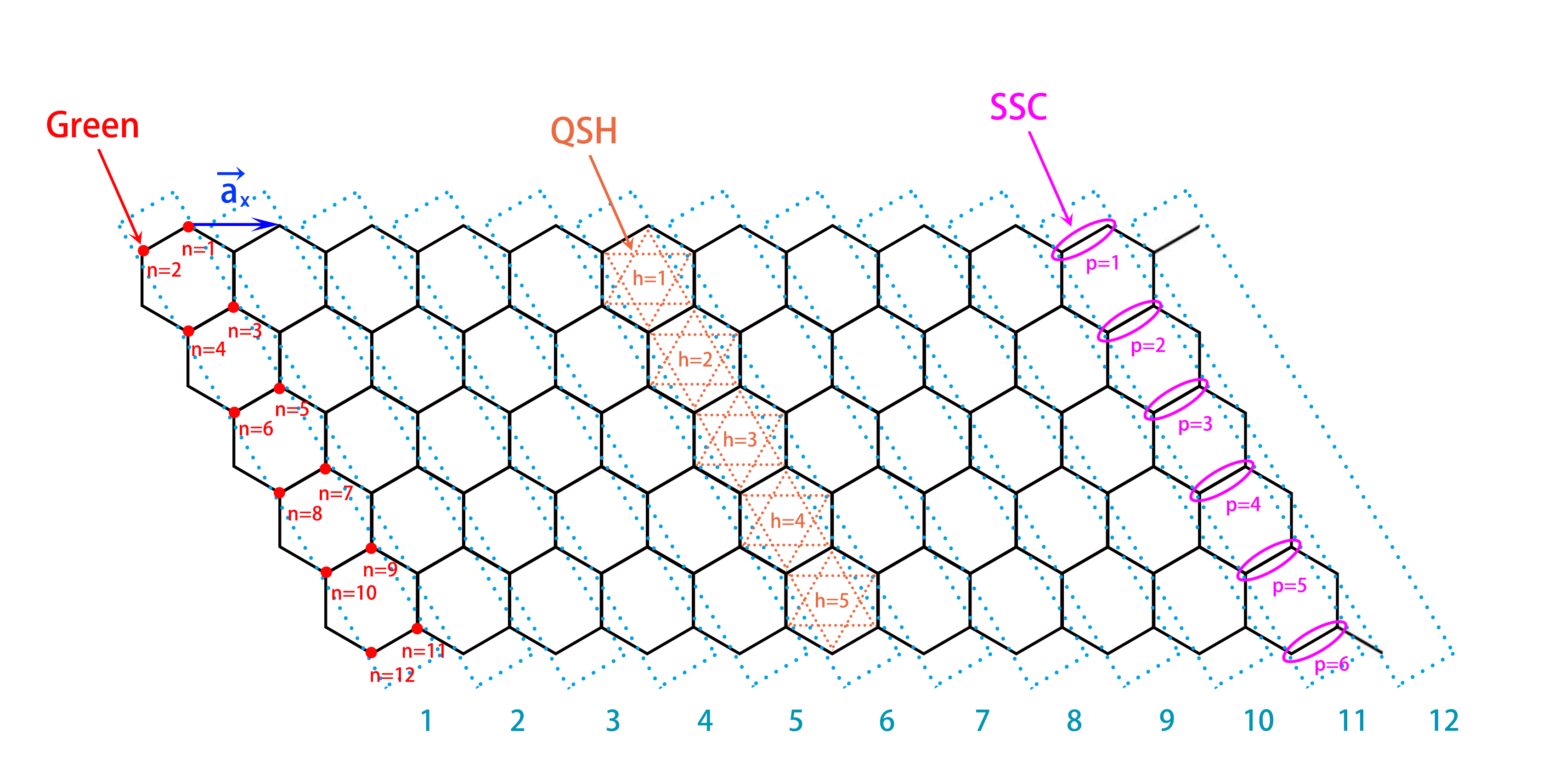}}
  \caption{\label{fig:lattice_diagram}
The honeycomb lattice with periodic boundary only in the $x$ direction with $N_{orb}=L_x=12$ orbitals per unit cell,
The model defines a unit cell with lattice vector $ \bm{a}_x$ and orbitals $N_{orb}$.
The boundary realizes a zigzag chain. }
\end{figure}

Fig.~\ref{fig:lattice_diagram} illustrates the geometry that we have used in this work. Here we consider a honeycomb lattice with  $L_x=12$ unit cells, each with $N_{orb}=L_x$ orbitals. This effectively  corresponds to a one dimensional lattice
with lattice vector  $\bm{a}_x$.
We use $p=1,2,\cdots,N_{orb/2}$ to label the bonds  defining the local
spin, charge and s-wave pairing operators, $h=1,2,\cdots,N_{orb/2}-1$ to label the hexagons for defining the QSH local operator, and $n=1,2,\cdots,N_{orb}$ to label the  orbitals.

We now provide the definition of correlation functions for the QSH, the s-wave pairing, the spin and the charge operators.
The QSH operator  is  defined on a hexagon $h$ in unit cell $r$ as:
\begin{equation}
  \label{Eq.QSH}
\hat{\boldsymbol{O}}^{\text{QSH}}_{  {r},  {h},\langle\langle\ve{\delta}_h,\ve{\delta}'_h\rangle\rangle}  =
i\hat{{c}}^{\dagger}_{  {r},  {h},\ve{\delta}_h} \ve{\sigma}
\hat{{c}}^{\phantom\dagger}_{ {r},  {h}, \ve{\delta}'_h}   +  \text{H.c.},
\end{equation}
where  $\ve{\delta}_h$ and $\ve{\delta}'_h$ are the legs of  a  next-nearest neighbour bond $\langle\langle\ve{\delta}_h,\ve{\delta}'_h\rangle\rangle $ in the hexagon.
The associated  time-displaced correlation functions reads
\begin{eqnarray}\label{Eq:CF1}
&&S^{\text{QSH}}_{h,\langle\langle\ve{\delta}_h,\ve{\delta}_h',\rangle\rangle, h' \langle\langle\ve{\delta}_{h'}'',\ve{\delta}_{h'}'''\rangle\rangle}(k_x, \tau)= \nonumber \\
&&\frac{1}{L_x}\sum_{ {r,r'} } \langle \hat{\ve{O}}^{\text{QSH}}_{ {r,h},\langle\langle\ve{\delta}_h,\ve{\delta}_h'\rangle\rangle}(\tau)\cdot \hat{\ve{O}}^{\text{QSH}}_ { {r',h'},\langle\langle\ve{\delta}_{h'}'',\ve{\delta}_{h'}'''\rangle\rangle}(0) \rangle
e^{i k_x ({r- r'}) }\;.  \nonumber \\
&&
\end{eqnarray}
The s-wave pairing operator is defined as
\begin{eqnarray}
  \label{Eq.SSC}
  \hat{O}^{\text{SSC}}_{ {r,p}, \ve{n}_p }&=&   \frac{1}{2} \left( \hat{c}^{\dagger}_{ {r,p},\ve{n}_p,\uparrow}  \hat{c}^{\dagger}_{ {r,p}, \ve{n}_p,\downarrow}+\text{H.c.}\right)
 \end{eqnarray}

Here $\ve{n}_p $ runs over the 2 orbitals corresponding to a unit cell $r$ and bond $p$ shown in Fig.~\ref{fig:lattice_diagram},
and the associated time-displaced correlation functions for the above-mentioned three operators are described as
\begin{eqnarray}\label{Eq:CF2}
&&S^O_{  p, \ve{n}_{p},  p', \ve{n'}_{p'} }(k_x, \tau)=\nonumber \\
&& \frac{1}{L_x}\sum_{ {r,r'} } \langle \hat{ \ve{O}}_{ {r}, p, \ve{n}_p }(\tau) \cdot \hat{\ve{O}}_ { {r'}, p',  \ve{n'}_{p'} }(0) \rangle
e^{i k_x ( {r- r'}) }  \;.
\end{eqnarray}
For  each correlation function we define the susceptibility by integrating over the imaginary time  $\tau$:
\begin{equation}
  \chi^O_{p,\ve{n}_p,p',\ve{n'}_{p'}}(k_x) = \int_0^{\beta} d\tau S^O_{p,\ve{n}_p,p',\ve{n'}_{p'}}(k_x,\tau)
\end{equation}
Choosing the bonds $p$ and $p'$ on the boundary allows to define edge or bulk correlation functions.  Having  fixed $p$ and $p'$ on the boundary we  still have a  $2 \times 2$ matrix that we diagonalize, and concentrate on the largest eigenvalue. A similar procedure is used the QSH  susceptibility.

\section{Localization of edge states}
\begin{figure}[h]
 \centering
\centerline{ \includegraphics[width=0.49\textwidth]{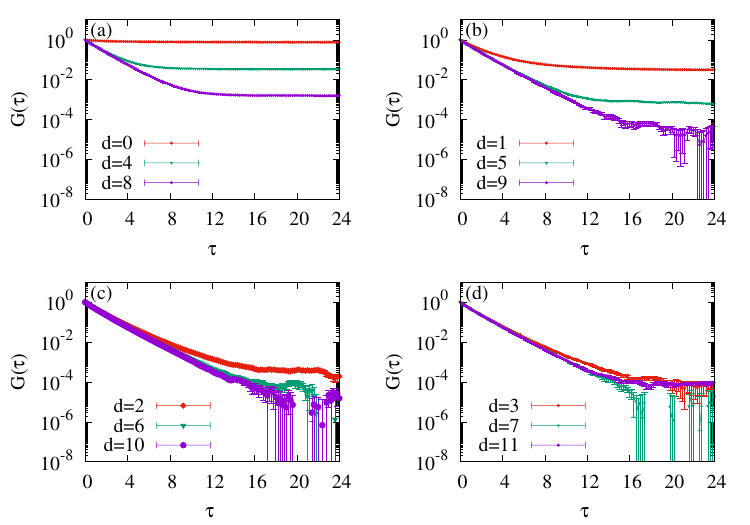}}
  \caption{\label{fig:SPG_lam0.026} Imaginary time Green function at $k=\pi$ as a function of distance $d$ from the boundary, for the
  t-$\lambda$ model in the QSH state at $\lambda=0.026$. }
  \label{fig:localization_QSH}
\end{figure}

\begin{figure}[h]
 \centering
\centerline{ \includegraphics[width=0.49\textwidth]{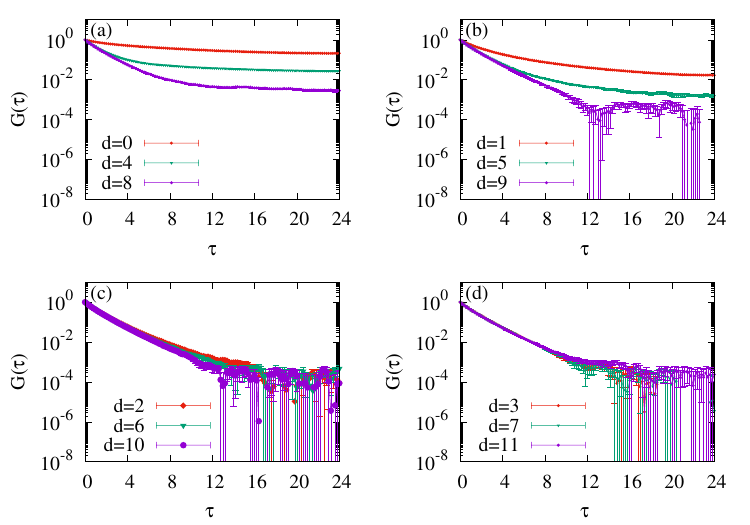}}
  \caption{\label{fig:SPG_lam0.0332} Imagninary time Green function as a function of distance $d$ from the boundary for the
  t-$\lambda$ model at the DQCP $\lambda=0.0332$. }
  \label{fig:localization_DQCP}
\end{figure}

To investigate the localization of edge states we compute the imaginary time  single particle Green function  as a function
from the distance $d$ to the boundary in the  QSH phase (Fig.~\ref{fig:SPG_lam0.026}) and at the DQCP (Fig.~\ref{fig:SPG_lam0.0332}).
In the QSH phase, the charge gap is finite, and it is expected that the edge states are localized.  We hence expect the imaginary time Green function at say $\tau = \beta/2$ to decay exponentially as a function of the distance $d$ to the boundary.
This is indeed what we observe in Fig.~\ref{fig:SPG_lam0.026}.   We note that there is an oscillation of period 4 in the decay such that we have considered four distinct panels for clarity.

Amazingly, the localization of the edge states is still present at the DQCP, associated with the finite single particle gap inside the bulk.  This can be seen from Fig.~\ref{fig:localization_DQCP}, where the exponential decay of Green's function does not differ quantitatively from the case of bulk QSH phase (see Fig.~\ref{fig:localization_QSH}).

\end{document}